\documentstyle{mn}

\newif\ifAMStwofonts

\input psfig
\def\simlt{\lower.5ex\hbox{$\; \buildrel < \over \sim \;$}}
\def\simgt{\lower.5ex\hbox{$\; \buildrel > \over \sim \;$}}

\def\etal{\rm et al.}
\def\msol{M$_{\sun}$}

\title[The Intermediate Age Brown Dwarf LP\,944-20]
      {The Intermediate Age Brown Dwarf LP\,944-20\thanks{
Based on observations made at the European Southern Observatory 3.6-m telescope, La Silla, Chile.}}

\author[C.G. Tinney]
       {C.G. Tinney \\
     Anglo-Australian Observatory, PO Box 296, Epping. N.S.W. 2121. Australia. 
      {\tt cgt@aaoepp.aao.gov.au}\\
       }

\date{Accepted ---.
      Received ---;
      in original form 10 November 1997}

\pagerange{\pageref{firstpage}--\pageref{lastpage}}
\pubyear{1995}

\begin{document}

\maketitle

\label{firstpage}

\begin{abstract}
Observations are presented which show that Li\,I\,$\lambda$6708 is detected 
with equivalent width of 0.53$\pm$0.05\AA\ in the proper-motion object LP\,944-20
(which is also known as BRI\,0337-3535). This Li detection implies that
LP\,944-20 is a brown dwarf with mass between 0.057 and 0.063\msol\ and age between 475
and 650\,Myr, making it the first brown dwarf to have its mass and age precisely determined.
\end{abstract}


\section{Introduction}

LP\,944-20 may possibly rank as one of the most overlooked objects 
in the astronomical literature.
It was first catalogued by Luyten \& Kowal in 1975, and remained unobserved for
a further 15 years until it was independently re-discovered as a possible very late star
by the Automated Plate Measuring machine (APM) colour QSO survey of Irwin, McMahon \& Hazard (1991).
This survey, alongside its main aim of finding high redshift quasars, turned up a number of
extremely red objects, which were subsequently determined to be M dwarfs 
(Kirkpatrick, Henry \& Irwin 1997; Irwin, McMahon \& Reid 1991). 
The re-discovery of LP\,944-20 lead to it
being given the new name of BRI\,0337-3535, by which it was referred to in the 
literature for several years before its orginal identification was pointed out by 
Kirkpatrick \etal\ (1997). Photometry of LP\,944-20 has been published by 
Kirkpatrick \etal\ (1997) and Tinney (1996) -- the latter work also presenting
a triginometric parallax. The data presented here were obtained in February 1994 as part of a 
study of the velocities of the lowest mass stars (Tinney \& Reid
1998).

\section{Observations \& Analysis}

High resolution spectra of a sample of very low-mass (VLM) stars were obtained 
using the Cassegrain Echelle Spectrograph (CASPEC) on the ESO 3.6m telescope
on the night of 1994 Feb 8 (UT). The resolution obtained was 16\,km/s (0.35\AA\ at 6708\AA), at
a dispersion of 9\,km/s per pixel. The wavelength range covered was
6400\AA\ to 9100\AA, though redwards of 8040\AA\ the wavelength coverage is not
complete and suffers inter-order gaps. The data were reduced using the
FIGARO data reduction system (Shortridge 1993) in a standard manner; the steps
included bias subtraction, flat-fielding, cosmic ray cleaning, echelle order straightening, 
object extraction, wavelength calibration, blaze function correction and flux calibration.
The data processing and analysis will be described in more detail by Tinney \& Reid (1998).

Figure 1 shows the region around the Li\,I\,$\lambda$6708 line in three of the objects
observed; LP\,944-20, and the cannonical late M-dwarfs LHS\,2065 and LHS\,2397a.
Table 1 shows the observational details relating to these three objects, together
with estimates of their spectral type from Kirkpatrick \etal\ (1995, 1997), and
absolute magnitudes based on trigonometric parallaxes.  Also shown in Table
1 are the heliocentric radial velocities derived for thes objects by cross-correlation
with the spectrum of VB8, which is assumed to have a radial velocity of $+$14.5$\pm$1\,km/s
(Tinney \& Reid 1998).

%
\begin{figure}
 \centerline{
                 \psfig{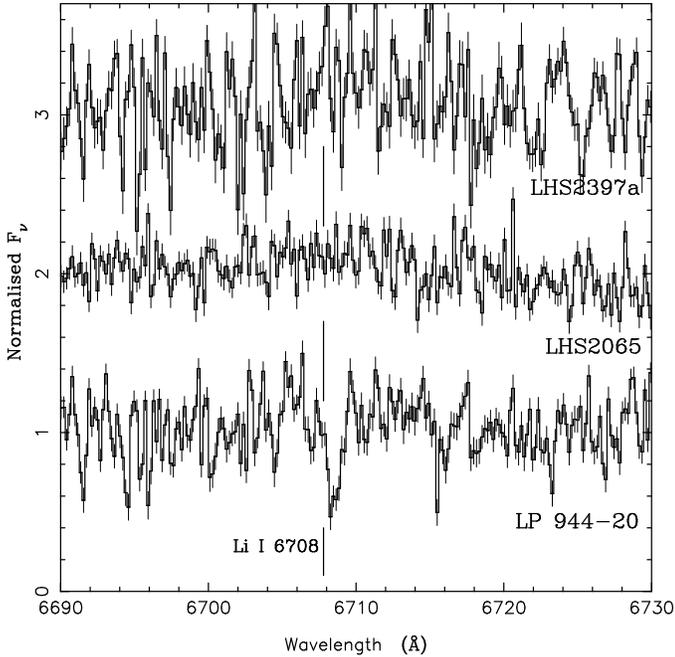}
            }
\caption{Figure 1 : CASPEC spectra of LP\,944-20, LHS2065 and LHS2397a in the region
of the  Li\,I\,$\lambda$6708 line. Each spectrum has been normalised at 6660-6700\,\AA,
and offset in unit steps for clarity. The error bars shown are photon counting
uncertainties, which have been propogated through the reduction. There are no atmospheric
absorption or night sky emission features in the neighbourhood of the Li\,I\,$\lambda$6708 line.}
\end{figure}

%
%
\begin{table*}
\begin{center}
  \center
  \caption{CASPEC Observations -- February 1994}
  \begin{tabular}{lcccccccr}
 Name                     &  Date       & Time &   I    &  I--K& M$_{K}$        & M$_{Bol}$$^a$  & Spectral & V$_{hel}$ \\
                          &   (UT)      & (s)  &        &      &                &                &  Type$^b$&(km/s)    \\[10pt]
LP\,944-20 (BRI\,0337-3535)& 1994 Feb 8 & 3600 &  14.16 & 4.58 & 11.10$\pm$0.06$^c$ & 14.32$\pm$0.07 & $>$M9V&10.0$\pm$2.0 \\
LHS2065     (LP\,666-9)    & 1994 Feb 7 & 5400 &  14.44 & 4.46 & 10.33$\pm$0.05$^d$ & 13.53$\pm$0.05 &   M9V & 9.2$\pm$2.0\\
LHS2397a    (LP\,732-94)   & 1994 Feb 7 & 5400 &  15.05 & 4.26 & 10.02$\pm$0.06$^e$ & 13.22$\pm$0.07 &   M8V &34.1$\pm$2.0\\
  \end{tabular}
\raggedright
\noindent
\vskip 10pt
$a$ -- M$_{Bol}$ derived using BC$_K$ of Tinney, Mould \& Reid 1993. $b$ -- Kirkpatrick \etal 1997, Kirkpatrick \etal\ 1995.
$c$ -- Tinney 1996. $d$ -- Monet \etal\ 1992; Tinney, Mould \& Reid 1993. $e$ -- Monet \etal\ 1992; Tinney 1996.\\
\end{center}
\end{table*}

The obvious
feature in Fig. 1 is the presence of an absorption near the wavelength of
Li\,I\,$\lambda$6708 line in LP\,944-20 -- and the absence of such a feature in LHS\,2065 and 
LHS\,2397a. Both of the latter are known to have depleted their Li, with
Mart\'\i n, Rebolo \& Magazz\`u (1994) placing 1.65-$\sigma$ upper limits on their
Li\,I\,$\lambda$6708 equivalent widths (EWs) of 0.15 and 0.7\,\AA\
(respectively). Because of the strength of molecular absorption in
the spectra of VLM stars, it is almost impossible to fit a true continuum for 
the purpose of estimating absorption EWs. Therefore, the
spectrum of LHS\,2065 was used as a template continuum. Its spectral type of M9V 
(Kirkpatrick, Henry \& Simons 1995)
makes it a close match in type to the $\ge$M9V of LP\,944-20. The LHS\,2065 
spectrum was smoothed with a 1.5 pixel 
Gaussian (to decrease its shot noise), redshifted
to the apparent radial velocity of the LP\,944-20, and suitably normalised.
The resulting EW = 0.53$\pm$0.05\,\AA, where the uncertainty is
determined by the photon counting errors shown in Fig 1. A fit to the position
of the Li\,I\,$\lambda$6708 absorption indicates an apparent radial velocity
of $+$30.4$\pm$4\,km/s, or a heliocentric radial velocity of $+$12.6$\pm$4\,km/s -- 
consistent with the radial velocity obtained by cross-correlation on the 
entire spectrum (cf. Table 1).
In the observing configuration Li\,I
falls in {\em two} orders of the echellogram -- the line was clearly
detected in both orders, though with lower significance than seen in Fig.1, 
which shows the merged orders.

H$\alpha$ is detected in emission -- but only weakly, with an emission EW = 1.2$\pm$0.5\AA, 
a full width at half maximum of $\approx$50\,km/s and at a heliocentric velocity of 0$\pm$10\,km/s.

\section{Discussion}

Kirkpatrick \etal\ (1997) have asigned a spectral type of $\ge$M9V to LP\,944-20,
making it similar to, or slightly cooler than, LHS2924 (M9V), and giving 
an approximate effective temperature of 2200K (Jones \etal\ 1994). The coldest 
available Li models 
are those of Pavlenko \etal\ (1995) at T$_{eff}$=2500 and 2000K. They show that
for EW = 0.53\,\AA\ the curve of growth is insensitive
to temperature, and implies a Li abundance of n(Li) $\approx$ 0.0 (where n(H)=12).
Given the likely initial abundance of n(Li)$\sim$2-3, this means that 
although Li is certainly present, it has been depleted by 2-3 dex.

\begin{figure*}
 \centerline{
                 \psfig{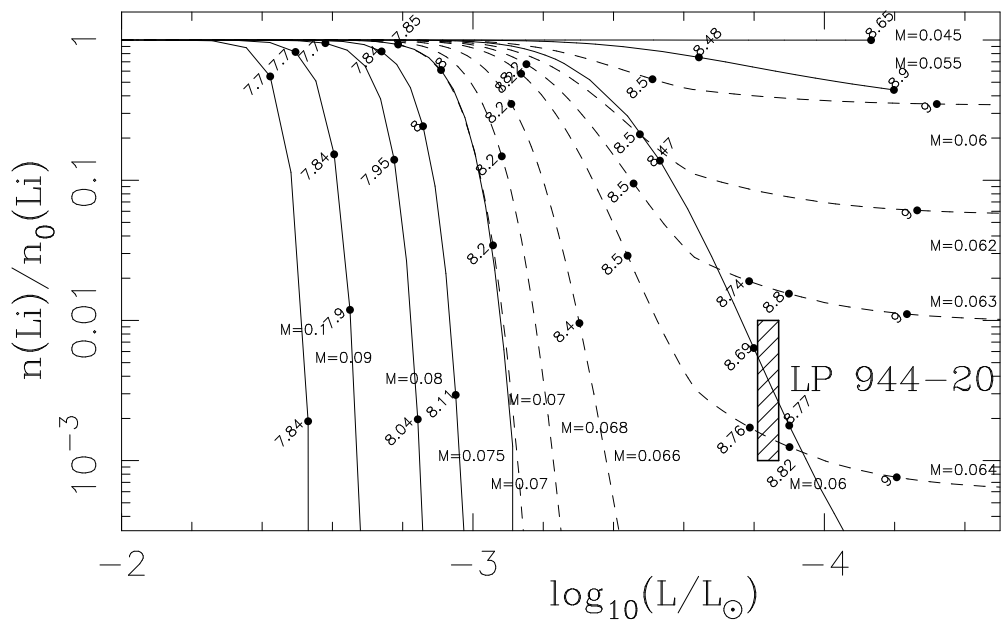}
            }
\caption{Figure 2 : Lithium depletion versus luminosity models in the brown dwarf regime,
due to Nelson \etal\ (dashed lines) and Chabrier \etal\ (solid lines). The masses of
each model sequence is indicated, as samples logarithmic ages. The shaded region
shows the location of the LP\,944-20 error box in this plane.}
\end{figure*}

The known luminosity of LP\,944-20 and its Li depletion enable us to constrain
its mass and age. Assuming M$_{bol\sun}$=4.72, the M$_{bol}$ of LP\,944-20
implies log(L$/$L$_{\sun}$)=$-$3.84$\pm$0.03. Interiors models including the effects of Li 
depletion have been
constructed by Chabrier, Baraffe \& Plez (1996), Ushomirsky \etal\ (1997),  
Nelson, Rappaport \& Chiang (1994), and D'Antona \& Mazzitelli (1994). The models of 
Ushomirsky \etal\ however, cannot be directly compared with our data, because they
present results in terms of T$_{eff}$ instead of luminosity. In Fig 2, we plot the
relevant models in log(n(Li)/n$_0$(Li)) versus luminosity from Chabrier \etal\  and Nelson \etal. 
The shaded region represents 
the error box imposed by our luminosity and lithium depletion results.
%
%
Chabrier \& Baraffe (1997) state that the Li destruction produced by the older
generation of Nelson \etal\ (1994) models is less efficient than that in the newer 
models due to ``the grey approximation, which yields larger $L$ and $T_{eff}$ and thus
central densities, favouring the onset of degeneracy .... and also smaller Graboske
\etal\ screening factors''. We therefore consider the Chabrier \etal\ (1996) models
to be the best currently available, but also show the Nelson \etal\ models to give
some idea of the sort of model-dependent uncertainties which may be present in our
derived mass and age for LP\,944-20, and because they lie on a denser mass grid. 

Bearing this in mind we conservatively estimate by interpolating between the 
Chabrier \etal\ models that the age of  
LP\,944-20 lies somewhere in the range
475--650\,Myr and that its mass lies in the range 0.057-0.063\,\msol.
This makes it considerably 
older than the brown dwarfs known in the Pleiades (Basri, Marcy \& Graham 1996; Rebolo \etal\ 1996), 
but considerably  younger than a typical old disk star.
This is by far the most precise mass and age estimate yet derived for a brown dwarf.
This is due to: (1) the precise luminosity determination; and (2) the
location of LP\,944-20 in a part of the n(Li)/L diagram, where there is little degeneracy
between age and mass.

The H$\alpha$ emission detected is at a very low level. It is, for example, significantly
lower than the EW = 3-10\AA\ seen in similar mass Pleiades-age objects
(Zapatero-Osorio \etal\ 1997; Hodgkin, Jameson \& Steele 1995; Stauffer \etal\ 1994)
implying an age $\simgt$150\,Myr. In fact the level of emission is more like
that seen in the brown dwarf DENIS-P\,J1228.2-1547,
which has age $\simlt$1\,Gyr (Tinney, Delfosse \& Thackrah 1997). The level
of chromospheric activity therefore argues for an age in the 200-1000\,Myr range. 

\section{Conclusion}

The discovery of a $\approx$500\,Myr brown dwarf which (other than having Li in
its spectrum) looks just like a M9V very low-mass star, adds considerable
support to the interpretation of the object 296A (Thackrah \etal\ 1997)
as a brown dwarf. In spite of a clear Li detection, the brown dwarf nature of 
this object has been questioned because of its very early M6V spectal type.
The discovery of the brown dwarf nature of LP\,944-20, however,
suggests a sequence of $\sim 0.06$\msol\ objects can be constructed: 
from very faint and probably old ($\sim$1\,Gyr) brown dwarfs like 
DENIS-P\,J1228.2-1547 and Kelu-1 (Tinney \etal\ 1997; Ruiz, Leggett \& Allard 1997); 
through somewhat younger (500-1000\,Myr) brown dwarfs like LP\,944-20;
to roughly Pleiades age brown dwarfs like 296A.

LP\,944-20 is the first brown dwarf for which tight (ie $\pm10\%$) constraints can
be placed on mass and age -- largely due to the knowledge of a precise luminosity, and
despite being only able to guess at
the initial Li abundance (which means that even with better data and modelling it is unlikely
the error bar on the Li depletion will fall much below 0.5 dex). It is clear that parallax
programs targetted at objects like Kelu-1, DENIS-P\,J1228.2-1547 and 296A are essential
to a more complete understanding of the mass and age of field brown dwarfs.

\subsection*{Acknowledgments}

I an enormously grateful to G.Chabrier and L.Nelson for providing digital
versions of their respective Li depletion models at short notice. I would also like to thank 
the technical and astronomical support staff at 
ESO La Silla for their assistance throughout 
this observing program, and to thank S.Ryan and J.Spyromilio
for helpful discussions on echelle data reduction.

\end{document}